\documentstyle{article}[12pt]
\def\ra{\rightarrow}

\def\es{\emptyset}
\def\ify{\infty}
\def\ov{\overline}
\def\sbs{\subset}
\def\ts{\times}

\def\a{\alpha}

\def\lb{\lambda}
\def\Om{\Omega}
\def\s{\sigma}
\def\Si{\Sigma}
\def\ve{\varepsilon}

\def\build#1_#2^#3{\mathrel{
\mathop{\kern 0pt#1}\limits_{#2}^{#3}}}

\font\tenbb=msbm10
\font\sevenbb=msbm7
\font\fivebb=msbm5
\newfam\bbfam
\textfont\bbfam=\tenbb
\scriptfont\bbfam=\sevenbb
\scriptscriptfont\bbfam=\fivebb
\def\bb{\fam\bbfam}

\def\Rb{{\bb R}}

\def\sgn{\mathop{\rm sgn}\nolimits}
\def\Aut{\mathop{\rm Aut}\nolimits}
\def\diag{\mathop{\rm diag}\nolimits}

\begin{document}

{\centerline {\bf Graphic requirements for multistationarity }}

\bigskip

{\centerline { Christophe Soul\'e }}

\medskip

{\centerline { CNRS et Institut des Hautes \'Etudes  Scientifiques}}

\bigskip
{\centerline { 2-06-2003 }}
\bigskip

\bigskip

IH\'ES 35 Route de Chartres 91440  Bures-sur-Yvette France

Tel:33 1 60 92 66 00 
  
  Fax: 33 1 60 92 66 09 

Email: soule@ihes.fr 

  \bigskip

\bigskip

\newpage

\centerline {\bf Abstract}
\bigskip

We discuss  properties which must be satisfied
by  a genetic network in order for it
to allow differentiation.

These conditions are expressed as follows in mathematical terms.
Let $F$ be a differentiable mapping from a finite dimensional real vector space to
itself. The signs of the entries of the Jacobian matrix of $F$ at a given
point $a$ define  an interaction graph, i.e. a finite oriented finite graph
$G(a)$ where each edge is equipped with a sign.
Ren\'e Thomas conjectured
twenty years ago that, if $F$ has at least two non degenerate zeroes,
there exists $a$ such that $G(a)$ contains a positive circuit.
Different authors proved this in special cases,
and we give here a general proof of the conjecture.
In particular, we get this way a necessary
condition for genetic networks to lead to multistationarity,
and therefore to differentiation.

We use for our proof the mathematical literature on global univalence,
and we show how to derive from it
 several variants of Thomas' rule,
some of which had been anticipated by Kaufman and Thomas.

\bigskip

\newpage 

\bigskip

When studying interactions in a system of biochemical
compounds, it is quite rare that one obtains quantitative results.
One can show that, in a given tissue,
 (the product of) a gene $A$ is an activator
(or a repressor) of the expression of a gene $B$, but the
strength of this interaction, the concentrations and their
kinetics are usually unknown. The resulting information is
essentially summarized by an {\it interaction graph} $G$, by
which one means a finite oriented graph together with a sign for
every edge. The vertices correspond to the members of the
network, and there is a positive (resp. negative) edge from $A$
to $B$ when $A$ activates (resp. represses) the synthesis of
$B$. Note that there can exist both a positive and a negative
edge from $A$ to $B$, since $A$ can activate $B$ at some
concentration and repress it at some other.

\smallskip

These interaction graphs can be quite complicated. It is
therefore very desirable to find ways to use them to restrict
the possible behaviour of the network they represent. In this
paper, we shall address the question of when the network is
susceptible to have several stationary states. A beautiful
conjecture of R.~Thomas \cite{T1} asserts that a {\it necessary}
condition for multistationarity  is that $G$ has an (oriented)
circuit $C$ which is {\it positive}, i.e. such that the product
of the signs of the edges of $C$ is positive,
at least in part of phase space. This was already
proved in several cases \cite{PMO} \cite{S} \cite{G} \cite{CD}, and
we shall present a proof in the general case (Theorem~1).

\smallskip

To formulate the question in mathematical terms, we choose
a continuous
model. Let $n \geq 1$
be an integer, and let $F$ be a differentiable map from $\Rb^n$
to itself. The evolution of a network of $n$ compounds can be
modelled by the differential system
$$
\frac{dx}{dt} = F(x) \, ,
$$
where $x$ is a differential path in $\Rb^n$: the components of
$x(t)$ are the different concentrations at time $t$. To ask
whether this system has several stationary states amounts to ask
whether $F$ has several zeroes. Given any $a \in \Rb^n$, 
the  interaction graph $G(a)$ is
defined from the  signs of the partial derivatives
$(\partial f_i / \partial x_j) (a)$ of the components of $F =
(f_i)$ at $a$ (see 2.1 below for a precise
definition) : a positive (resp. negative) sign indicates that $j$
is an activator (resp. a repressor) of $i$.
The precise formulation of Thomas' rule is then the following:
if $F$ has at least two (non-degenerate) zeroes in $ \Rb^n$, 
there exists $a \in \Rb^n$ such that $G(a)$
contains a positive circuit.

\smallskip

The main remark leading to a proof of 
this assertion is the following.
If $F$ has several zeroes, it cannot be univalent (i.e. one to
one). Therefore, if we know {\it sufficient} conditions for $F$
to be univalent, their negation will give necessary conditions
for $F$ to have several zeroes. And we might then deduce
properties of $G(a)$ from these necessary conditions.

\smallskip

It turns out that finding sufficient conditions for $F$ to be
univalent is a classical issue in mathematical economy, when one wants
to know that the factor prices are uniquely determined
by the prices for goods.
Several results
have been obtained with this application in mind. G.~Gale and
H.~Nikaido \cite{GN} gave an elegant criterion of univalence in
terms of the Jacobian matrix $J(x) = (\partial f_i / \partial
x_j)(x)$: assume that, for every $x \in \Rb^n$, the principal minors
(resp. the determinant) of $J(x)$ are nonnegative (resp. is positive);
then $F$ is univalent. As we shall see, this result leads precisely
to a proof of Thomas' conjecture (Theorem~1).

\smallskip

 Following the same line of arguments, we can also apply variants of
the theorem of Gale-Nikaido, which are discussed for instance in the
book of T.~Partha\-sarathy \cite{P}. A motivation for doing so is
that one may want to restrict the domain of $F$ to be the closed
positive quadrant (since a concentration cannot be negative). In
general, $F$ will be defined on a product $\Om$ of $n$ intervals,
open or not. However, in this generality,
there exist counter-examples to the condition of Thomas (see 3.5). But,
 when $F$ has at least two zeroes,
we can still find properties of  its interaction graphs
(Theorem 5), as a consequence of a  univalence theorem of
Gale-Nikaido (\cite{GN} , \cite{P}). Furthermore, it is often
the case that, when defining the interaction graph of several
biochemical compounds, the degradation of each of them is not taken
into account. We show in Theorem~6 that this makes Thomas' rule
valid in full generality.

\smallskip

Another refinement of the theorem of Gale-Nikaido, due to
C.B.~Garcia and W.I.~Zangwill \cite{GZ}, applies to the case where
$\Om$ is closed and bounded. This leads to Theorem~7 which, in turn,
for general $\Om$, gives some information on the location of the
zeroes of $F$ (Theorem~8).

\smallskip

The economist P.A.~Samuelson imagined a stronger univalence
criterion that those above. It is not true for general $F$, but
L.A.~Campbell proved it when $\Om = \Rb^n$ and $F$ is {\it
algebraic}, e.g. when each component of $F$ is the quotient of two
real polynomials. In Theorem~9 we translate his result in terms of
properties of the interaction graph.

\smallskip

One can expect more graphical requirements for multistationarity. One
of them is a conjecture of M.~Kaufman  and  R.~Thomas
\cite{TK}, which
is stronger than the original Thomas' conjecture. We have been
unable to prove this assertion, but we obtained some evidence for it
in Theorems 3, 4 and 5.

\smallskip

The paper is organized as follows. The first section gives
definitions about graphs, matrices and determinants. It shows basic
lemmas, which are standard knowledge in the literature on
interaction graphs. In section two we define the interaction graphs
$G(a)$ and we state the conjectures of Thomas and Kaufman-Thomas. In
the third paragraph we prove the conjecture of Thomas and we give
some (counter)examples. In section four we discuss the
conjecture of Kaufman-Thomas. Next, we give results when the domain
$\Om$ is not necessarily open, and we refine them in section six.
Finally, we discuss the case of an algebraic map in the last
section.

\smallskip

The literature on the question studied here is scattered in several
different journals, and it has been crucial for me to receive papers
from different authors. I want to thank for that J.~Aracena,
O.~Cinquin, J.~Demongeot, J.L.~Gouz\'e, M.S.~Gowda, M.~Kaufman,
D.~Thi\'efry and R.~Thomas. I am also grateful to the participants
of the ``S\'eminaire d'initiation \`a la g\'enomique fonctionnelle"
of IH\'ES, and especially to
F.~K\'ep\`es, who explained  the importance of genetic
networks, and first mentioned to me the rule of Thomas. Finally, I am
extremely thankful to M.~Kaufman and R.~Thomas for several
discussions, where they patiently and carefully explained their
ideas, provided examples and encouraged me to do this
work.

 \bigskip

    \section{Graphs and matrices}

   \subsection{ \ } An {\it interaction graph} $G = (V,E,\sgn)$
is a finite oriented graph $(V,E)$ together with a {\it sign}
map $\sgn : E \ra \{ \pm 1 \}$. In other words, $V$ (the
vertices) and $E$ (the edges) are two finite sets and each
edge $e \in E$ has an origin $o(e) \in V$ and an endpoint
$t(e) \in V$ (it may happen that $o(e) = t(e)$).
 \smallskip

A {\it  circuit} in $G$ is a sequence of edges $e_1
, \ldots , e_k$ such that $o(e_{i+1}) = t(e_i)$ for all
$i = 1,\ldots , k-1$ and $t(e_k) = o(e_1)$.
\smallskip

A {\it hooping} is a collection $C = \{ C_1 , \ldots , C_k
\}$ of circuits such that, for all $i \ne j$, $C_i$ and $C_j$ do
not have a common vertex. A circuit is thus a special case of
hooping. We let $V(C) = \build\coprod_{i=1}^{k} V(C_i)$ be
the (unordered) set of vertices of $C$. This set $V(C)$ will
also be called the {\it support} of $C$. A hooping is
called {\it Hamiltonian} when its set of vertices is maximal,
i.e.
$V(C) = V$. Note that hoopings are called "generalized circuits",
or "g-circuits" in \cite{ED}, and
Hamiltonian ones are called
"full circuits" in   \cite{TK}.

\smallskip

The {\it sign} of a  circuit $C$ is
$$
{\rm sgn} \, (C) = \prod_{e \in C} {\rm sgn} \, (e) \in
\{ \pm 1 \} \, .
$$
When ${\rm sgn} (C) = + 1$ (resp. $-1$) we say that
$C$ is positive (resp. negative).
The sign of a hooping $C$ is
\begin{equation}
\label{eq1}
{\rm sgn} \, (C) = (-1)^{p+1} \, ,
\end{equation}
where $p$ is the number of positive  circuits in
$C$ (cf. \cite{ED}).

           \smallskip

Given any subset $I \sbs V$ we let $\tau_I \, G$ the interaction
graph obtained from $G$ by changing the sign of every edge $e
\in E$ such that $t(e) \in I$. Given any permutation $\s \in
\Aut (V)$ of the vertices, we let $\s G$ be the interaction
 graph obtained from
$G$ by replacing each edge $j \ra i$ by an edge $j \ra \s (i)$,
with the same sign.

   \subsection{ \ } Let $n \geq 1$ be an integer and $A =
(a_{ij})$ an $n$ by $n$ real matrix. We can attach to
$A$ an interaction graph $G$ as follows. The set of
vertices of $G$ is $\{ 1 , \ldots , n \}$. There is an
edge $e$ with $o(e) = j$ and $t(e) = i$ if and only if
$a_{ij} \ne 0$. The sign of $e$ is the sign of $a_{ij}$.

\smallskip

Given any subset $I \sbs \{ 1 , \ldots , n \}$, the {\it
principal minor} of $A$ with support $I$ is the real
number ${\rm det} \, (A_I)$, where $A_I$ is the square
matrix $(a_{ij})_{i,j\in I}$. By definition
\begin{equation}
\label{eq2}
{\rm det} \, (A_I) = \sum_{\s \in \Si_I} {\ve (\s)}
\prod_{i \in I} a_{i\s(i)} \, ,
\end{equation}
where $\Si_I$ is the group of permutations of $I$ and
$\ve(\s)$ is the signature of $\s$
(Recall that $\ve$ is defined by the equalities $\ve (\s \s') =
\ve (\s) \, \ve (\s')$ for all $\s , \s' \in \Si_I$ and $\ve
(\s) = -1$ when $\s \in \Si_I$ is a transposition).

For any $\s \in \Si_I$ we let
\begin{equation}
\label{eq3}
a(\s) = {\ve(\s)} \prod_{i \in I} a_{i \s (i)}
\end{equation}
so that
$$
{\rm det} \, (A_I) = \sum_{\s \in \Si_I} a(\s) \, .
$$
    When $a (\s) \ne 0$, we let ${\rm sgn} \, (a(\s)) = \pm 1$
be its sign.

      \smallskip

Let $D = \diag (d_1 , \ldots , d_n)$ be a diagonal $n$ by $n$
real matrix and $I \sbs \Si$ any subset. It follows from the
definition (2) that
\begin{equation}
\label{eq4bis}
\det ((A+D)_I) = \sum_{J \sbs I} \det (A_J) \prod_{i \in I-J}
d_i \, .
\end{equation}

                                         \smallskip

Given $I \sbs V$ any subset, we let $\tau_I \, A$ be the matrix
obtained from $A$ by replacing $a_{ij}$ by $-a_{ij}$ whenever $i
\in I$. Given any $\s \in \Si_n = \Aut (V)$ we let $\s A$ be the
product of $A$ with the permutation matrix defined by $\s$.
Clearly
$$
G(\tau_I \, A) = \tau_I \, G(A)
$$
and
$$
G(\s A) = \s G(A) \, .$$

\subsection{ \ }     We keep the notation of the
preceeding paragraph.
    Note that, given any permutation $\s \in \Si_I$,
     there is a unique decomposition
$$
I = I_1 \amalg I_2 \amalg \ldots \amalg I_k
$$
of $I$ into a disjoint union of nonempty subsets such that
the restriction $\s_{\a}$ of $\s$ to $I_{\a}$ is a {\it
cyclic permutation} for all $\a = 1, \ldots , k$. Let
$C(\s_{\a})$ be the  circuit of $G$ with edges $(i ,
\s_{\a} (i))$, $i \in I_{\a}$ (note that $a_{i \s_{\a}
(i)} \ne 0$ since $a(\s) \ne 0$). We denote by $C(\s)$ the
hooping of $G$ which is the disjoint union of
the  circuits $C(\s_{\a})$, $\a = 1,\ldots ,k$.

 When $X$ is a finite set, we let $\# \,
(X)$ be its cardinality. The following lemma is due to J.Eisenfeld and
C.DeLisi (\cite{ED}, Appendix, Lemma 2), and is
 probably at the origin of the
definition (1).

\medskip

\noindent {\bf Lemma 1.} {\it Let $I \sbs \{ 1 , \ldots ,
n \}$ be any subset and let $\s \in \Si_I$ be such that $a(\s)
\ne 0$. Then the following identity holds
$$
{\rm sgn} \, (C(\s)) = {\rm sgn} \, (a(\s)) \, (-1)^{\#
(I) + 1} \, .
$$
}

\medskip

\noindent {\bf Proof.} Since $\ve (\s) = \build\sum_{\a =
1}^{k} \ve (\s_{\a})$ and $I = \ \build\coprod_{\a}^{}
I_{\a}$, we get from (\ref{eq3}) that
\begin{equation}
\label{eq4}
a(\s) = \prod_{\a} a (\s_{\a}) \, .
\end{equation}
For each $\a = 1,\ldots ,k$, since $\s_{\a}$ is cyclic we
have
$$
\ve (\s_{\a}) = (-1)^{\# (I_{\a}) + 1} \, .
$$
The circuit $C(\s_{\a})$ is positive if and only if
$\build\prod_{i \in I_{\a}}^{} a_{i \s_{\a} (i)}$ is
positive, in which case we get from (\ref{eq3}) that
\begin{equation}
\label{eq5}
{\rm sgn} \, (a(\s_{\a})) = (-1)^{\# (I_{\a}) + 1} \, .
\end{equation}
When $C(\s_{\a})$ is negative we get
\begin{equation}
\label{eq6}
{\rm sgn} \, (a(\s_{\a})) = (-1)^{\# (I_{\a})} \, .
\end{equation}
Therefore, by (\ref{eq4}), (\ref{eq5}), (\ref{eq6}), we
get
$$
{\rm sgn} \, (a(\s)) = (-1)^p \, (-1)^{\build\sum_{\a}^{}
\# (I_{\a})} = (-1)^p \, (-1)^{\# (I)} \, ,
$$
where $p$ is the number of positive circuits in
$C(\s)$.
Since, by (\ref{eq1}), ${\rm sgn} \, (C(\s)) = (-1)^{p+1}$,
the lemma follows.

                 \subsection{ \ } From Lemma~1 we get the following:

\medskip

\noindent {\bf Lemma 2.} {\it Let $A$ be an $n$ by $n$ real
matrix and let $I \sbs  \{ 1 , \ldots , n \}$ be any subset.

\smallskip

\noindent {\rm i)} Assume that $\det (-A_I)$ is negative (resp.
positive). Then there exists a positive (resp. negative)
hooping in $G(A)$ with support $I$.

\smallskip

\noindent {\rm ii)} Assume $\det (-A_I) = 0$. Then either there
exist two hoopings in $G(A)$ with opposite signs and
support $I$, or there is no hooping in $G(A)$ with
support $I$.}

\medskip

\noindent {\bf Proof.} From (\ref{eq2}) we get
$$
\det (-A_I) = (-1)^{\# (I)} \det (A_I) \, .
$$
Therefore, by Lemma~1,
\begin{equation}
\label{eq8}
\det (-A_I) = - \sum_{\s \in \Si_I} \sgn (C(\s)) \vert a(\s)
\vert \, .
\end{equation}
Assume $\det (-A_I)$ is negative (resp. positive). Then, by
(\ref{eq8}), there exists $\s \in \Si_I$ such that $C(\s)$ is
positive (resp. negative). This proves i).

\noindent If $\det (-A_I) = 0$, either there exist two summands
with opposite signs on the right hand side of (\ref{eq8}), or
$a(\s)$ is zero
for all $\s \in \Si_I$. This proves ii), since every
hooping of $G(A)$ with support $I$ is of the form $C(\s)$
for some $\s \in \Si_I$.
                                \section{Conjectures on multistability}

\subsection{ \ }
Let $\Om_i$ be a nonempty interval in $\Rb$:
$$
\Om_i = \ ] a_i , b_i [ \, , \ [a_i , b_i[ \, , \ ]a_i , b_i] \
\hbox{or} \ [a_i , b_i] \, ,
$$
with $a_i \geq -\ify$ and $b_i \leq +\ify$. Denote by $\Om \sbs
\Rb^n$ the product $\Om = \build\prod_{i=1}^{n} \Om_i$.
 Consider a
map
$$
F : \Om \ra \Rb^n
$$
which is {\it differentiable}, i.e. such that, for each
$i,j \in \{ 1 , \ldots , n \}$ and any $a \in \Om$, the
$i$-th component $f_i$ of $F$ has a partial derivative
$\frac{\partial f_i}{\partial x_j} \, (a)$ at the point
$a$ and
$$
f_i (x) = f_i (a) + \sum_{j=1}^{n} \frac{\partial
f_i}{\partial x_j} \, (a) (x_j - a_j) + o (\Vert x-a
\Vert) \, ,
$$
where $\Vert x-a \Vert$ is the norm of $x-a$ and $o$ is
the Landau $o$-symbol (\noindent we could also use
 a weaker notion of differentiability, see
\cite{GR}). The map $F$ is called $C^1$ when every partial
derivative $\partial f_i / \partial x_j$ is continuous on $\Om$.

For any $a \in \Om$, the {\it Jacobian} of $F$ at $a$ is
the $n$ by $n$ real matrix
$$
J(a) = J(F)(a) = \left( \frac{\partial f_i}{\partial x_j} \,
(a) \right) \, .
$$

\smallskip

For any $a \in \Om$ we let
$$
G(a) = G(J(a))
$$
be the interaction graph of $J(a)$, defined as in 1.2. We also
let $G(F)$ be the interaction graph defined as follows. Its set
of vertices is $V = \{ 1 , \ldots , n \}$. Given $i$ and $j$ in
$V$, there is at most one positive (resp. negative) edge from
$j$ to $i$; it exists if and only if there is a positive (resp.
negative) path in $G(a)$ for some $a \in \Om$. In other words,
$G(F)$ is the ``superposition'' of all the interaction graphs $G(a)$.

\smallskip

Given any subset $I \sbs V$, we let $\tau_I \, F$ be the map
obtained from $F = (f_i)$ by replacing $f_i$ by $-f_i$ when $i
\in I$. Given $\s \in \Si_n$, we let $\s F = (f_{\s^{-1} (i)})$.
The interaction graphs of $\s \tau_I \, F$ are $\s \tau_I \,
G(a)$, $a \in \Om$, and $\s \tau_I \, G(F)$.

\subsection{ \ } We shall be interested in the set of
zeroes of $F$, i.e. the points $a \in \Om$ such that $F(a)
= 0$. They can be viewed as the {\it stationary states} of
the system of differential equations
$$
\frac{dx(t)}{dt} = F(x) \, ,
$$
where $x(t)$ is a differentiable mapping from a real
interval to $\Om$.

\smallskip

We say that a zero $a$ of $F$ is {\it nondegenerate} when
${\rm det} (J(a)) \ne 0$.

\subsection{Conjecture 1. (Thomas \cite{T1})} Assume that
$\Om$ is open and that $F$ has at
least two nondegenerate
zeroes in $\Om$. Then there exists $a \in \Om$
such that  $G(a)$ contains a {\it positive}  circuit.

                                         \medskip

\noindent {\bf Remark.} This conjecture has already been proved
in several cases. First, it is known to hold when the signs of the
entries in $J(a)$ are independent of $a \in \Om$ \cite{PMO}
\cite{G}; see also \cite{S} and the remark in 5.2 below. It was
also shown in \cite{CD} for stable stationary states when $\Om$
contains the positive quadrant and
$f_i (x) > 0$ whenever $x_i = 0$.

\subsection{Conjecture 2. (Kaufman-Thomas \cite{TK})}
Under the same assumption as Conjecture~1,

\smallskip

i) Either there exist $a \in \Om$
such that
$G(a)$  has a positive
Hamiltonian hooping
and  a negative
Hamiltonian hooping.

\smallskip

ii) Or  there is a cyclic
permutation $\s$ of a subset of
$\{ 1 , \ldots , n \}$
and there exist $a,b \in \Om$ 
 such that the  circuit
$C(\s)$ in $G(a)$ (resp. in $G(b)$) is positive (resp.
negative).

\bigskip

Note that Conjecture~2 for $F$ implies Conjecture~1 for
$F$. It is quite different though, since positive and negative circuits
play in it a symmetric role.
In case i) we shall say that
"$G$ has two Hamiltonian hoopings of opposite signs"
and in case ii)  we shall
 say that ``$G$
 has an ambiguity''. In the latter case,
$G(F)$ contains two circuits with the same ordered set
of vertices and opposite signs (but ii) is stronger 
than that statement).

\section{A proof of Thomas' Conjecture~1}

\subsection{ \ }
\noindent {\bf Theorem 1.} {\it Assume $\Om$ is open and $F$ has 
at least two nondegenerate zeroes in $\Om$. Then, for every $I 
\sbs V$ and every $\s \in \Si_n$, there exists
$a \in \Om$ such that $\s \tau_I \, G(a)$ has a
positive circuit. In particular, Thomas' Conjecture 1 holds 
true.}

 \bigskip

 Since $\s \tau_I \, G(x) = G (\s \tau_I J(x))$
and since $\s \tau_I (F)$ satisfies the hypotheses of Theorem~1 
if and only if $-F$ does, we just have to check that,
 for some $a \in \Om$, the interaction graph
 $G(-J(a))$ has
a positive circuit. According to Lemma~2 i), it will be enough 
to show that, for some $a \in \Om$, a principal minor of $-J(a)$
is negative. In other words:

\medskip

\noindent {\bf Theorem 1'.} {\it Assume that, for every $a \in 
\Om$, all the principal minors of $-J(a)$ are nonnegative. Then 
$-F$ can have at most one nondegenerate zero.}

\subsection{ \ } For any positive real number $\lb$ let
$$
\phi_{\lb} (x) = -F(x) + \lb x \, .
$$
Since
$$
J (\phi_{\lb}) = -J(F) + \diag (\lb , \ldots , \lb) \, ,
$$
it follows from (\ref{eq4bis}) that, under the hypotheses of 
Theorem~1', each principal minor of $J (\phi_{\lb})$ is positive 
on $\Om$. According to Gale and Nikaido, \cite{GN} Theorem 4, 
this implies that $\phi_{\lb}$ is univalent. The following 
proposition ends the proof of Theorem~1':

\medskip

\noindent {\bf Proposition 1.} {\it Let $\phi = \Om \ra \Rb^n$ 
be a differentiable map defined on an open set $\Om \sbs \Rb^n$. 
Assume that, for all $\lb > 0$, the map
$$
\phi_{\lb} (x) = \phi (x) + \lb x
$$
is univalent. Then $\phi$ can have at most one nondegenerate 
zero in $\Om$.}

\subsection{ \ } The proof of Proposition~1 proceeds as 
\cite{GN} Theorem 4' and \cite{P}, IV, Theorem 4, p.~35. Assume 
$a$ and $b$ are two nondegenerate zeroes of $\phi$ in $\Om$. 
According to \cite{AH}, XII, \S~2.9, p.~477, we can choose open 
neighborhoods $U_a$ and $U_b$ of $a$ and $b$ respectively such 
that $\ov U_a \cap \ov U_b = \es$, $\ov U_a \sbs \Om$, $\ov U_b 
\sbs \Om$, $a$ (resp. $b$) is the unique zero of $\phi$ in $\ov 
U_a$ (resp. $\ov U_b$), and the degrees $\deg (\phi , \ov U_a , 
a)$ and $\deg (\phi , \ov U_b , 0)$ are equal to $\pm 1$. 
Arguing as in \cite{P} (or \cite{GN}), loc.cit., we get
$$
1 = \deg (\phi , \ov U_a \cup \ov U_b , 0) = \deg (\phi , \ov 
U_a , 0) + \deg (\phi , \ov U_b , 0) \, ,
$$
hence a contradiction. \hfill q.e.d.

\subsection{ \ } The restriction to
nondegenerate zeroes of $F$ in Theorem~1 
is necessary. For 
example, if $n=2$, $\Om = \Rb^2$ and 
$$
F(x,y) = (-xy^2 , -y)
$$
we have
$$
-JF(x,y) = \left( \matrix{
y^2 & 2xy \cr
0 & 1 \cr} \right) \, .
$$
Clearly $G$ has no positive circuit. However $F(x,0) = 0$ 
for any $x \in \Rb$.

\subsection{ \ } It is also essential that $\Om$ be open. Let 
$\Om$ be the set of $(x,y) \in \Rb^2$ such that $x \geq 0$ and 
$y \geq 0$. Consider the map $F : \Om \ra \Rb^2$ defined by
$$
F(x,y) = ((y-2)^2 - x^2 - 1 , 4x - 4xy) \, .
$$
We get
$$
-JF (x,y) = \left( \matrix{2x &4-2y \cr 2y-4 &2x \cr} \right) \, 
.
$$
Therefore, by (\ref{eq8}), $G(F)$ does not have any positive 
circuit. On the other hand both $(0,3)$ and $(0,1)$ are 
nondegenerate zeroes of $F$ in $\Om$. We shall discuss the case 
of an arbitrary $\Om$ in sections five and six below.

\section{On the conjecture of Kaufman-Thomas}

\subsection{ \ } We first assume that $n=2$. Let $\Om \sbs 
\Rb^2$ be as in 2.1 and open, with coordinates
 $x$ and $y$. For any $h : \Om
\ra \Rb$ we write $h \equiv 0$ to mean that $h(a) = 0$ for every 
$a \in \Om$. If $h$ is differentiable, we let $h_x$ (resp. 
$h_y$) be its partial derivative with respect to the first 
(resp. second) variable.

Let 
$$
F = (f,g) : \Om \ra \Rb^2
$$
be a differentiable mapping.

\medskip

\noindent {\bf Theorem 2.} {\it Assume that $F$ 
has at least two nondegenerate
zeroes in 
$\Om$. Then, one of the following conditions holds:}
\begin{itemize}
\item[i)] {\it $G$ has two Hamiltonian hoopings
 of opposite 
signs;}
\item[ii)] {\it $G$ has an ambiguity;}
\item[iii)] {\it $f_x \, g_y \equiv 0$ but $f_x \not\equiv 
0$ and $g_y \not\equiv 0$.}
\end{itemize}

\medskip

\noindent {\bf Proof.} Assume that $G$ has no ambiguity, and
that its Hamiltonian  hoopings have all the same sign. If, in 
addition, $f_x \, g_y \not\equiv 0$, since the sign of 
$f_x (a)$ (resp. $g_y (a)$) is independent of $a$, we can 
multiply $f$ and $g$ by $\pm 1$ to get to the case where 
$f_x (a) \geq 0$ and $g_y (a) \geq 0$ for all $a \in \Om$. 
It follows that $f_x (a) \, g_y (a) \geq 0$ and, since all 
Hamiltonian hoopings have the same sign, $f_y (a) \, g_x (a) \leq 
0$ (by Lemma~1). Therefore Theorem~1' applies to $F$ and we
conclude that $F$ has at most one nondegenerate zero.

\smallskip

Assume now that $g_y \equiv 0$. After multiplying $f$ and 
$g$ by $\pm 1$ we can assume that $f_x \geq 0$ and $f_y \, 
g_x \leq 0$. Once again, Theorem~1' implies that $F$ has at
most one nondegerate zero. q.e.d.

\subsection{ \ } With an additional assumption, 
Conjecture~2 is true for all $n \geq 2$:

\medskip

\noindent {\bf Theorem 3.} {\it Let $F : \Om \ra \Rb^n$ be
a differentiable mapping such that $\Om$ is open and
$F$ has at least two nondegenerate zeroes. Then,
one of the following conditions holds:}
\begin{itemize}
\item[i)] {\it $G$ has two Hamiltonian hoopings with opposite signs;}
\item[ii)] {\it $G$ has an ambiguity;}
\item[iii)] {\it for any point $a \in \Om$ there exists $i 
\in \{ 1 , \ldots , n \}$ such that $G(a)$ does not 
contain an edge from $i$ to itself.}
\end{itemize}

\medskip

\noindent {\bf Proof.} Assume that all Hamiltonian hoopings
 in $G$
have the same sign, that $G$ has no ambiguity and that,
for some $a \in \Om$, and for any $i \in \{ 1 , \ldots , 
n\}$, there is an edge in $G(a)$ from $i$ to itself. The 
last condition means that all diagonal entries in $J(a)$ 
are nonzero. After multiplying each component of $F$ by 
$\pm 1$ we can assume that all diagonal entries of $J(a)$ 
are positive. Let $C$ be any hooping in
$J(a)$, and let $I$ be its set of vertices.
 The disjoint union of
$C$ with all the circuits $i \ra i$, $i \in \{ 1 , 
\ldots , n\} - I$, is a Hamiltonian hooping of $G(a)$. Its sign 
must be the sign of the Hamiltonian hooping which is the disjoint 
union of
all the positive circuits $i \ra i$, $i \in \{ 1 , \ldots , 
n\}$, namely $(-1)^{n+1}$. We conclude that
$$
{\rm sgn} \, (C) = (-1)^{\# (I) } \, .
$$
By Lemma~1, this implies that $C = C(\s)$ with ${\rm sgn} 
(a(\s)) = + 1$, $\s \in \Si_I$. Since $G$ has no ambiguity,
for any $\s \in \Si_I$ we have ${\rm sgn} \, (a(\s)) \geq
0$ in $\Om$. Therefore, for any $x \in \Om$, all the 
principal minors of $J(x)$ are nonnegative.
 Applying Theorem~1', we
conclude that $F$ has at most one nondegenerate zero.

\subsection{ \ } Concerning the graph $G(F)$ (see
2.1) we can prove the following: 

\medskip

\noindent {\bf Theorem 4.} {\it Let $F : \Om \ra \Rb^n$ be 
a differentiable mapping such that $\Om$ is open and
$F$ has at least two nondegenerate zeroes. Then
$G(F)$ has two Hamiltonian hoopings with opposite signs.}

\medskip

\noindent {\bf Proof.} We first remark that we can find 
$a \in \Om$ and $\s \in \Si_n$ such that none of the diagonal entries 
in $J(\s F)(a)$ is zero. Indeed these entries are, by definition,
$ \frac{\partial f_{\s^{-1}(j)}}{\partial x_j} \,
(a) $, $j=1, \ldots , n$, and, when $a$ is a non degenerate zero
of $F$, it follows from (1) that, for some $\s \in \Si_n$,
$$ \prod_{i=1}^n\frac{\partial f_i}{\partial x_{\s (i)}} \,
(a) \neq 0.$$
This proves the claim.

We may then choose $I$ such that all the diagonal entries 
in $J(\tau_I \, \s F)(a)$ are negative. In other words,
for each $i \in \{ 1 , \ldots , n \}$, the interaction
graph
$\tau_I \,\s G(a)$ contains a negative edge
$i \ra i$. On the other hand, by Theorem 1, there exists
$b \in  \Om$ such that $\tau_I \,\s G(b)$
has  a positive circuit $C$. In $\tau_I \,\s G(F)$
the disjoint union of $C$ with the negative edges
$i \ra i, i \notin V(C)$, is a positive Hamiltonian hooping,
when the union of all the negative edges
 $i \ra i , i \in \{ 1 , \ldots , n \}$, 
is a negative one. 

It is thus enough to show that, if $\tau_I G(F)$ or
$\s G(F)$ has two Hamiltonian hoopings with opposite signs,
the same is true for $G(F)$. This is clear for $\tau_I G(F)$ 
and, for $\s G(F)$,  it follows from Lemma 3 below.

\subsection{ \ } 

\medskip

\noindent {\bf Lemma 3.} {\it Let $G$ be any interaction
graph, $\s \in {\rm Aut}(V)$ a permutation of its vertices,
and $C$ an Hamiltonian hooping in $G$.
The image of $C$ in the interaction
graph  $\s G$ is then an Hamiltonian hooping with 
sign $\ve (\s) {\rm sgn}(G)$.}

\medskip

\noindent {\bf Proof.} Recall from 1.1 that $\s G$
is obtained from
$G$ by replacing each edge $j \ra i$ by an edge $j \ra \s (i)$,
with the same sign. As a collection of edges,
$C$ has a well defined image $\s C$ in $\s G$.
To check Lemma 3 we may assume that $\s$ is the transposition of
two vertices $i$ and $j$ (transpositions span ${\rm Aut}(V)$).

Assume first that $i$ and $j$ are in the same circuit $C_1$ of $C$.
Then all the circuits in $C$ other than $C_1$ are fixed by $\s$.
The image of $C_1$ consists of two disjoints circuits $D_1$
and $D_2$. More precisely, if the vertices of
$C_1$ are $1 \, 2 \ldots i \ldots j \ldots m$
(as we can assume), we get
$$ \s C_1 = D_1 \coprod D_2 
$$
where the sequence of vertices in $D_1$ (resp. $D_2$)
is $1 \, 2 \ldots i-1 \, j \, j+1 \ldots k \, 1$
(resp. $ i \, i+1 \ldots j-1 \, i$).
Furthermore, if $C_1$ has an even (resp. odd) number of negative 
edges, $D_1$ and $D_2$ will have the same (resp. a different)
number of edges modulo two.
From this it follows that $\s C$ is a
Hamiltonian hooping such that, with the definition  
(\ref{eq1}),
$${\rm sgn}(\s C) = - {\rm sgn}(C) ,$$
as was to be shown.

Note that we also have
$$  C_1 = \s D_1 \coprod \s D_2,$$
therefore, by exchanging the roles of $C$
and $\s C$, the previous discussion applies 
also to the case where $i$ and $j$ lie in two different circuits
of $C$.

\hfill q.e.d.

\section{The case of a domain which is not open}

\subsection{ \ } We keep the notation of 2.1, where $\Om$ is an 
arbitrary product of intervals and $F : \Om \ra \Rb^n$ is 
differentiable.

\medskip

\noindent {\bf Theorem 5.} 

{\it
 Assume that $F$ is not univalent. Then:

\noindent {\rm 1)} For 
every $\s \in \Si_n$,

{\rm i)} Either there exists $a \in \Om$ and $i \in V$ such that 
$\s G (a)$ does not contain any edge from $i$ to itself;

{\rm ii)} Or, for any subset $I \sbs V$,  there exists
$a \in \Om$ such that  $\tau_I \,\s G(a)$ has
a positive circuit.

\smallskip

\noindent {\rm 2)} When the condition 1) i) above 
is not satisfied,
$G(F)$ contains both a positive and a
negative Hamiltonian hooping.}

\medskip

\noindent {\bf Proof.} The map $F$ is univalent if 
and only if $\tau_I \, \s F$ is. 
 So, to prove 1), we can restrict our attention to $-F$. Note 
also that i) is equivalent to the assertion that there exists $I 
\sbs V$ such that $\s G (a)$ does not contain any hooping 
with support I. According  to Lemma~2 ii), if none of the 
conclusions in 1) is true, all the principal minors of the 
Jacobian matrix of $-F$ are positive on $\Om$. By the 
Gale-Nikaido theorem, \cite{GN} Theorem~4, this implies that
$-F$ is univalent. This proves 1).

\smallskip

To prove 2), by replacing $F$ by $\tau_I \, F$ for an 
appropriate choice of $I$, we can  assume that, for each 
vertex $i$ in $G(F)$, there is a negative edge  from $i$ to 
itself. Since $G(F)$ also 
contains a positive circuit by 1), we get, as in 4.3 above, 
that $G(F)$ contains both a positive and a
negative Hamiltonian hooping.
 
\medskip

\noindent {\bf Remark.}Since the condition 1) i) is often satisfied,
when $\Om$ is open  Theorem 5 is much weaker than Theorems 1 and 4.

\subsection{ \ } Here is a variant of Theorem~5.

\medskip

\noindent {\bf Theorem 6.} {\it Assume given, for every $a \in 
\Om$, a diagonal matrix $D(a)$ with positive entries. Assume $F$ 
is not univalent. Then, for some $a \in \Om$, the interaction 
graph
$$
H(a) = G(J(a) + D(a))
$$
has a positive circuit.}

\medskip

\noindent {\bf Proof.} Assume that, for any $a \in \Om$, none of 
the circuits of $H(a)$ is positive. We know from Lemma~2 i) that 
all the principal minors of $-J(a) - D(a)$ are nonnegative. From 
(\ref{eq4bis}) this implies that all the principal minors of 
$-J(a)$ are positive, and, again by \cite{GN} Theorem~4, $F$ 
must be univalent.

\medskip

\noindent {\bf Remark.} Theorem~6 applies to the situation 
considered for instance in \cite{T2} and \cite{KT}, where $-D(a)$ 
comes from "terms of decay", which are not taken into
account when drawing the interaction graph. It was proven by 
Snoussi \cite{S} when the signs of the entries of $J(a)$ are 
constant.

\section{On the location of stationary states}

\subsection{ \ } Assume that $\Om = \build\prod_{i=1}^{n} [a_i , 
b_i]$ is a closed bounded subset of $\Rb^n$. In that case, 
Garcia and Zangwill got a stronger result that Gale-Nikaido 
(\cite{GZ}, see also \cite{P}, V, Theorem~1, p.~41). For any $I 
\sbs V$ define $I^c = V - I$ and, for any $n$ by $n$ real 
matrix $A$, let
$$
m_I (A) = \det (A_{I^c})
$$
and
$$
m_i (A) = m_{\{ i \}} (A)
$$
for each $i \in V$. When $x$ lies in $\Om$, we write $m_I (x)$ 
for $m_I (J(x))$ and $m_i (x) = m_i (J(x))$. Define $I(x) \sbs 
V$ as the set of vertices $i$ such that $x_i = a_i$ or $x_i = 
b_i$. The result of Garcia-Zangwill is the following

\medskip

\noindent {\bf Theorem 7'.} {\it Assume $F$ is $C^1$ and that, 
for every $a \in \Om$, and every subset $I \sbs I(a)$,
$$
m_I (a) \prod_{i \in I} m_i (a) > 0 \, .
$$
Then $F$ is univalent.}

\medskip

In particular, when $a \in \, \build\Om_{}^{\circ} \, = 
\build\prod_{i=1}^{n} \, ]a_i , b_i [$, $I(a) = \es$ and the 
only assertion made is that $\det J(a) > 0$.

\noindent When $F$ is only differentiable, see \cite{GR}, 
p.~930, Remark.

\subsection{ \ } Theorem~7' implies the following refinement of 
Theorem~5:

\medskip

\noindent {\bf Theorem 7.} {\it Assume that $\Om$ is bounded and 
closed, and that $F : \Om \ra \Rb^n$ is $C^1$. If $F$ is not 
univalent, for every $\s \in \Si_n$, one of the following 
conditions holds true:

\smallskip

\noindent {\rm i)} There exists $a \in \Om$ and $I \sbs I(a)$ such 
that no hooping of $\s G(a)$ has support $I^c$, or $\s G(a)$ 
contains two hoopings with support $I^c$ and opposite signs.

\smallskip

\noindent {\rm ii)} There exists $a \in \partial \Om = \Om \, - 
\build\Om_{}^{\circ}$, $I \sbs I(a)$ and hoopings $C_I$ 
and $C_i$, for each $i \in I$, in $\s G (a)$ such that $\# (I) 
\geq 2$, the support of $C_I$ (resp. $C_i$) is $I^c$ (resp. $\{ 
i \}^c$), and
$$
\sgn (C_I) \prod_{i \in I} \sgn (C_i) > 0 \, .
$$
Furthermore, either $\# (I)$ is even or there exist $b \in 
\partial \Om$, $I' \sbs I(b)$, $C_{I'}$, $C_j$, $j \in I'$, with 
similar properties as above and
$$
\sgn (C_{I'}) \prod_{j \in I'} \sgn (C_j) < 0 \, .
$$
}

\noindent {\bf Proof.} Again, it is enough to treat the case $\s 
= 1$. When $F$ is not univalent we know from Theorem~7' that 
there is $a \in \Om$ and $I \sbs I(a)$ such that
$$
m_I (a) \prod_{i \in I} m_i (a) \leq 0 \, .
$$
Assume $m_I (a) = 0$. Then, by Lemma~2 ii), the statement i) 
must hold.

\noindent Assume now, that, for all $J \sbs I(a)$, $m_J (a) \ne 
0$. Since
$$
m_I (a) \prod_{i \in I} m_i (a) < 0
$$
we must have $\# (I) \geq 2$. Furthermore, for every vertex $k 
\in V$, if we replace $F$ by $\tau_k \, F = \tau_{\{ k \}}\, F $ the quantity $m_I
(a)$ is multiplied by $+1$ (resp. $-1$) if $k \in I$ (resp. $k 
\in I^c$). Therefore, in all cases, $m_I (a) \build\prod_{i \in 
I}^{} m_i (a)$ gets multiplied by $(-1)^{\# (I)}$. It is thus
invariant if and only if $\# (I)$ is even. If this number
is odd, we apply the 
same discussion to $-F$ and we get $b$ and $I'$ such that
$$
m_{I'} (b) \prod_{j \in I'} m_j (b) > 0 \, .
$$
Using (\ref{eq8}), the statement i) follows.

\subsection{ \ } Let $\Om = \build\prod_{i=1}^{n} \Om_i$ be an 
arbitrary product of intervals as in 2.1 and let $F : \Om \ra 
\Rb^n$ be a $C^1$ map. From Theorem~7 one gets some information 
on where two zeroes of $F$ can be:

\medskip

\noindent {\bf Theorem 8.} {\it Fix $\s \in \Si_n$. Assume that 
there exist two points $a$ and $b$ in $\Om$ such that $F(a) = 
F(b)$. Assume furthermore that, when $x$ lies in 
$\build\Om_{}^{\circ}$, all hoopings of $\s G (x)$ are 
negative. Then there exists $x \in \Om$ such that, when $x_i \in 
\partial \Om_i$, the $i$-th coordinate of $a$ or $b$ is equal to 
$x_i$, and a subset $I \sbs I(x)$ such that no hooping in 
$\s G (x)$ has support equal to $I^c$.}

\medskip

\noindent {\bf Proof.} One can find a bounded closed product of 
intervals $\Om' \sbs \Om$ containing $a$ and $b$ and such that, 
whenever $x \in \Om'$ and $x_i \in \partial \Om_i$, the $i$-th 
coordinate of $a$ or $b$ is equal to $x_i$. Since the 
restriction of $F$ to $\Om'$ is not univalent, we can apply 
Theorem~7. We are not in case ii) because, $J(x)$ being 
continuous in $x$, for every $x \in \Om'$ all the 
hoopings of $G(x)$ are nonpositive.

\smallskip

Therefore Theorem~7 i) holds true for some $x \in \Om'$, hence 
the conclusion. 

\hfill q.e.d.

\section{The algebraic case}

Assume now that $\Om= \Rb^n$ and that $F$ is $C^1$ and {\it 
algebraic}, by which we mean that its graph $\{ (x,F(x))\} \sbs 
\Rb^n \ts \Rb^n$ is the set of zeroes of a family of real 
polynomials in $2n$ variables. This will be the case for 
instance when each component $f_i$ of $F$ is the quotient of two 
polynomials in $n$ variables. In that case, a result of Campbell 
\cite{C} leads to a stronger conclusion than Theorem~5 1).

\medskip

\noindent {\bf Theorem 9.} {\it Choose any ordering of $V$. Let 
$$
F : \Rb^n \ra \Rb^n
$$
be a $C^1$ map which is not univalent. Then we can choose $k 
\leq n$ such that, if $I$ consists of the first $k$ vertices,

\smallskip

\noindent {\rm i)} Either $G(F)$ has two hoopings of 
opposite signs and support equal to $I$;

\smallskip

\noindent {\rm ii)} Or there exists $a \in \Rb^n$ such that none 
of the hoopings of $G(a)$ has support equal to $I$.}

\medskip

\noindent {\bf Proof.} For any $k \leq n$ and $a \in \Rb^n$ we 
let
$$
d_k (a) = \det (J(a)_I) \, .
$$
 According to 
\cite{C}, if $d_k (a) > 0$ for all $k = 1, \ldots , n$, the map 
$F$ is univalent. Therefore, under our assumption, $d_k (a) 
\leq 0$ for some $a$ and some $k$. When $d_k (a) = 0$, either i) 
or ii) is true (by Lemma~2 ii)). Given $i \in V$, when we 
replace $F$ by $\tau_i \, F $, $d_k (a)$ 
gets multiplied by $-1$ (resp. $+1$) if $i \leq k$ (resp. $i > 
k$). So we can assume that, for every $I \sbs V$, there exist 
$k$ and $a$ such that $\tau_I \, d_k (a) < 0$ (with obvious 
notation). By replacing $F$ by $\tau_1 \, F$ we see that we can 
also assume that there exist $k'$ and $b$ such that $\tau_I \, 
d_{k'} (b) > 0$.

\smallskip

Unless i) or ii) holds, we can assume (using Lemma~2 again) that 
for every $I \sbs V$ and every $k \leq n$ the sign of $\tau_I \, 
d_k (x)$ does not depend on $x \in \Rb^n$. Let then $m$ be the 
maximum of all integers $p$ such that, if $i \leq k \leq p$, 
$d_k (x)$ and $d_1 (x)$ have the same sign. When $m < n$, we can 
increase $m$ by $m+1$ by replacing $F$ by $\tau_{m+1} \, F$. By 
repeating this process, we find $I$ such that $\tau_I \, d_k 
(x)$ has a fixed sign for every $k \leq n$ and every $x \in 
\Rb^n$. As we saw in the previous paragraph, this cannot happen. 
\hfill q.e.d.

\bigskip

Keywords :  interaction graph, multistationarity,
Jacobian matrix, global univalence.

\bigskip

\end{document}